\documentclass{INTERSPEECH2023}

\interspeechcameraready 

\usepackage{csquotes}
\usepackage{multirow}
\usepackage{rotating}

\title{Improving End-to-End Neural Diarization Using Conversational Summary Representations}
\name{Samuel J. Broughton, Lahiru Samarakoon}
\address{
  Fano Labs, Hong Kong SAR, China
}
\email{\{samuel.broughton,lahiru\}@fano.ai}

\begin{document}

\maketitle
 
\begin{abstract}
Speaker diarization is a task concerned with partitioning an audio recording by speaker identity. End-to-end neural diarization with encoder-decoder based attractor calculation (EEND-EDA) aims to solve this problem by directly outputting diarization results for a flexible number of speakers. Currently, the EDA module responsible for generating speaker-wise attractors is conditioned on zero vectors providing no relevant information to the network. In this work, we extend EEND-EDA by replacing the input zero vectors to the decoder with learned conversational summary representations. The updated EDA module sequentially generates speaker-wise attractors based on utterance-level information. We propose three methods to initialize the summary vector and conduct an investigation into varying input recording lengths. On a range of publicly available test sets, our model achieves an absolute DER performance improvement of 1.90 \% when compared to the baseline.
\end{abstract}

\section{Introduction}

Speaker Diarization is a task concerned with determining the number of speakers and their respective speech activities for a given input audio signal, often referred to as the problem of \textit{\enquote{who spoke when?}} \cite{ryant2020third}. A diarization system is considered to be robust if it can perform well when dealing with overlapping speech segments, long form audio and an arbitrary number of speakers across a range of acoustic domains \cite{chung2020spot, kinoshita2021advances}.

Traditional methods employed to achieve diarization have involved clustering-based approaches within a modular pipeline structure \cite{anguera2012speaker, park2022review}. Given an input audio signal, speaker active frames are detected with a voice activity detection module (VAD). Speaker embedding vectors are extracted from speaker active frames and clustered so that segments belonging to the same speaker can be labelled. The main disadvantage with clustering-based approaches is that they cannot handle overlapping speech, an inherent characteristic of real world conversational data \cite{chung2020spot, yu2022m2met}. Some methods have been designed to address this problem \cite{diez2018but, kinoshita2021integrating}, however, this substantially increases the complexity of the solution and inter-module dependencies.

End-to-end neural-based diarization systems directly solve the overlapping speech issue, simplify the overall design of the diarization pipeline and generally perform better than traditional approaches. End-to-end neural diarization (EEND) directly outputs diarization results by treating the task as a multi-label classification problem \cite{fujita2019end}. EEND-based models require permutation invariant training (PIT), as they currently predict diarization results without taking into consideration speaker order \cite{yu2017permutation}. An encoder-decoder attractor calculation module (EDA) can replace the classification head in EEND to flexibly determine the number of speakers in a given utterance \cite{horiguchi2020end, horiguchi2021encoder}. The EDA module is responsible for creating speaker-wise attractor representations for diarization results calculation and speaker existence prediction.

However, the LSTM-based EDA module has limitations. LSTM networks are prone to vanishing gradients when handling long sequences such as recordings used for training in diarization. This means for recordings containing a higher number of active speakers, the EDA module will struggle to produce well-separated attractor representations. To tackle this problem, the original authors shuffled input embeddings to the LSTM encoder \cite{horiguchi2021encoder}, and studied the use of both global and local attractor calculation \cite{horiguchi2021towards}. One key issue is that the EDA LSTM decoder is expected to sequentially generate speaker-wise attractor representations from only the last hidden and cell states of the LSTM encoder. The input zero vector to the decoder contains no relevant information to the network and is common to all recordings processed by the model. By estimating a more meaningful representation the model could perform better for recordings containing a higher number of active speakers.

% Even though EEND-EDA can handle a flexible number of speakers at inference, it is capped by that of the dataset it is trained on. To tackle this problem, the original authors presented various ways to generalize EEND-EDA for recordings containing more speakers \cite{horiguchi2021towards}. One key issue is the zero vector input to the LSTM decoder of the EDA module, which contains no relevant information to the network when generating speaker-wise attractor representations \cite{pan2022towards}. The only input features are the last hidden and cell states of the LSTM encoder used to initialize the hidden and cell states of the LSTM decoder. As a result, the model suffers at inference when an input recording contains higher numbers of active speakers.

%The diarization error rate (DER) measures the performance of a diarization system. DER is composed of missed speech (MS), false alarm (FA) and speaker error (SPK-ERR) metrics \cite{anguera2012speaker}. Currently there is no standard practice for reporting DER results. Often, works will report DER results across entire test datasets [CITE]. Sometimes this includes a breakdown of MS, FA and SPK-ERR [CITE]. Others will provide a breakdown of results split by the number of active speakers within a recording [CITE], showing that diarization becomes a harder task as the number of active speakers increases.

This work presents a feature-based approach to enhance attractor representations in the EDA module of EEND-EDA for better diarization performance in recordings containing a higher number of active speakers. Inspired by the special classification token in BERT \cite{devlin2018bert}, we introduce a summary vector to the first frame of the subsampled acoustic input feature sequence to learn a conversational summary representation. This learned summary replaces the input zero vector to the EDA module previously used, giving the LSTM decoder additional utterance-level information. Incorporating conversational summary representations to the EDA module achieves a lower diarization error rate (DER) when compared to the baseline, particularly for recordings with a higher number of active speakers. 

In Section \ref{sec:eend-eda} we revisit EEND-EDA, Section \ref{sec:summ-vec} outlines the usage of the summary vector, Section \ref{sec:data-experiments} explains the experimental conditions, Section \ref{sec:results} discusses the results, Section \ref{sec:conclusion} concludes the paper.

\section{EEND-EDA} \label{sec:eend-eda}

EEND is the current state-of-the-art end-to-end diarization architecture used by the community \cite{fujita2019end, kinoshita2021advances, khare2022asr, ueda2022eend, yu2022auxiliary}. An EDA module was introduced to allow EEND to handle an arbitrary number of speakers \cite{horiguchi2020end, horiguchi2021encoder, horiguchi2021hitachi, leung2021end, leung2021robust}.

The main training objective of EEND-based diarization is to learn a mapping function that outputs the likelihood of speech activities for multiple speakers given an input sequence $X \in \mathbb{R}^{D \times T}$. Here, $X$ is a log-scaled Mel-filterbank acoustic input feature sequence with dimension $D$ and length $T$. Output embedding $E \in \mathbb{R}^{D \times T}$ is produced by passing $X$ as input to a Transformer-based encoder \cite{vaswani2017attention, gulati2020conformer}.
Time shuffled embedding $E$ is input to an LSTM-based EDA module to predict speaker-wise attractors $A \in \mathbb{R}^{D \times (S + 1)}$, where $S$ is the number of speakers.

%Where $s \in \{1, ..., S\}$ and $t \in \{1, ..., T\}$ is the number of speakers and frame index, respectively.%

Diarization results are calculated by a sigmoid function operating on the element-wise multiplication of $A$ and $E$. This outputs posterior probabilities $(\textbf{p}_t)^T_{t=1}$ for the speech activities of multiple speakers across all frames. Where $\mathbf{p}_t := [p_{t, 1}, ..., p_{t, S}] \in (0, 1)^S$ are the posterior probabilities for $S$ speakers at frame $t$. These posteriors are without conditions to decide the order of the speakers. During inference, the scalar value of $p_{t, s}$ is used with a threshold value of $0.5$ to determine whether speaker $s$ should be labelled as active or not at frame $t$.

\textbf{Diarization Loss.} A permutation free objective is used to optimize the model by calculating the loss for all possible speaker assignments between diarization predictions and groundtruth labels $(\mathbf{y}_t)^T_{t=1}$ \cite{yu2017permutation}. Where $\mathbf{y}_t := [y_{t, 1}, ..., y_{t, S}] \in \{0, 1\}^S$ are the groundtruth labels for all speakers at frame $t$. Here,  $y_{s, t} = 1$ and $y_{s, t} = 0$ denotes whether speaker $s$ is active or not at $t$. The minimum loss is then taken for backpropagation and can be defined as:

\begin{equation}
    \label{eq:diar}
    \mathcal{L}_{diar} = \frac{1}{TS} \min_{i \text{ } \in \text{ perm}(1, ..., S)} \sum^T_{t=1} H(\mathbf{p}^i_t, \mathbf{y}_t),
\end{equation}

where perm$(1, ..., S)$ and $\mathbf{p}^i_t$ resemble the set of permutations for all speakers and the permuted posterior labels at frame $t$, respectively. Here, $H(\cdot, \cdot)$ is the binary cross entropy.

\textbf{Attractor Existence Loss.} The output hidden and cell states of the EDA LSTM encoder $h^{\text{enc}}$ are used to initialize the hidden and cell states of the EDA LSTM decoder $h^{\text{dec}}$. Speaker-wise attractors are generated from input zero vectors:

\begin{align}
    \label{eq:orig_lstmd}
    \textbf{h}^{\text{dec}}_{s}, \textbf{c}^{\text{dec}}_{s} &= h^{\text{dec}}(\textbf{0}, \textbf{h}^{\text{dec}}_{s-1}, \textbf{c}^{\text{dec}}_{s}) \qquad \qquad (s=1, ..., S),
\end{align}

where hidden state $\textbf{h}^{\text{dec}}_{s}$ corresponds to speaker $s$'s attractor. Attractor existence posterior probabilities are calculated by inputting $A$ to a fully connected layer followed by sigmoid activation. Attractor existence loss $\mathcal{L}_{exist}$ is calculated using equation (19) in \cite{horiguchi2021encoder}.

The \textbf{full training objective} of EEND-EDA can be defined as:

\begin{equation}
    \mathcal{L} = \mathcal{L}_{diar} + \alpha \mathcal{L}_{exist},
\end{equation}

where $\alpha$ is a hyperparameter.

\section{Summary Vector for EEND-EDA} \label{sec:summ-vec}

\begin{figure}[t!]
    \centering
	\includegraphics[width=\linewidth]{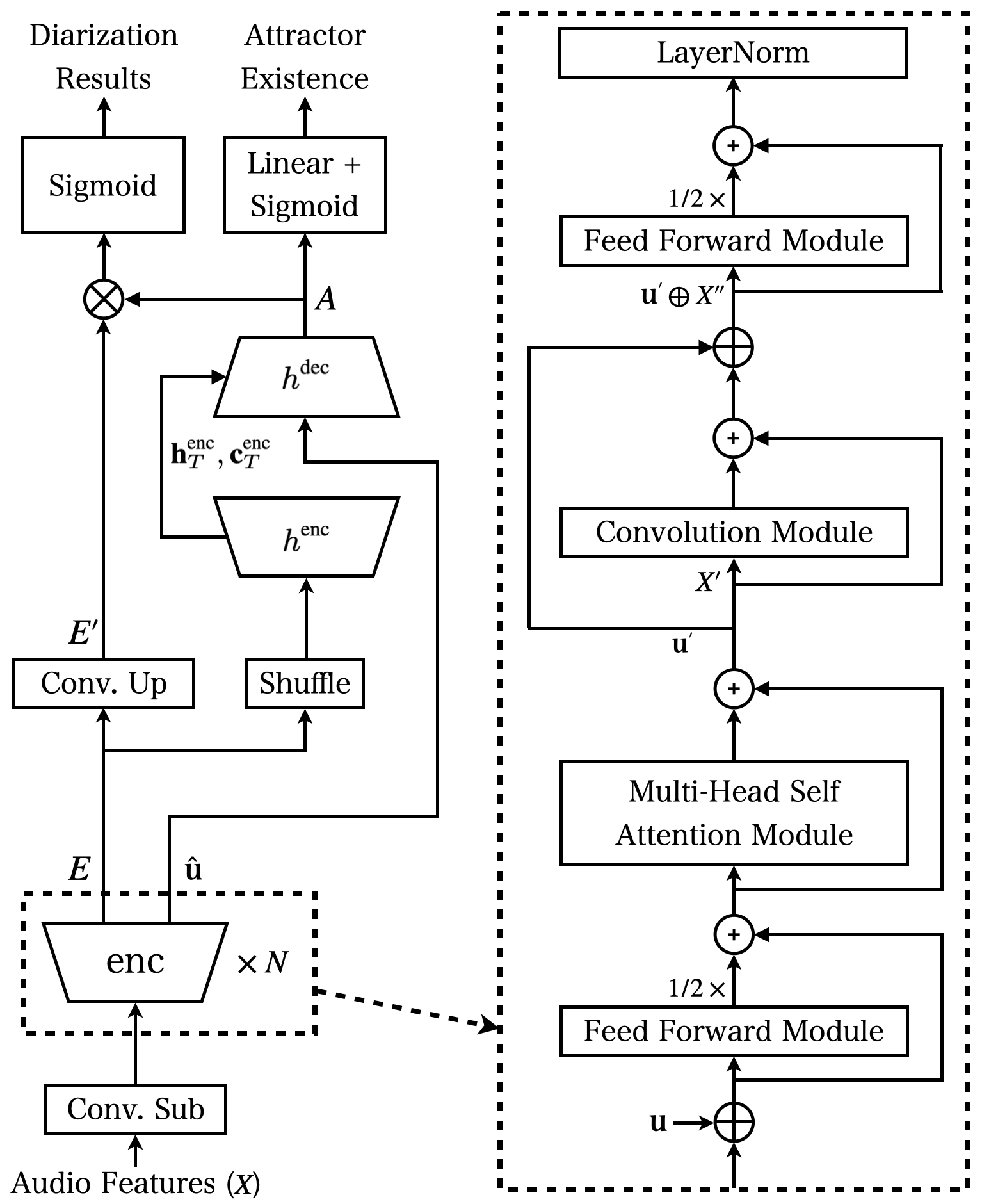}
	\caption{Proposed network architecture for EEND-EDA with special summary vector representation. The network is similar to EEND-EDA with convolutional subsampling and upsampling \cite{leung2021robust}. The dashed box on the right shows the modification to Conformer encoder when used with a special summary vector representation.} 
	\label{fig:network}
\end{figure}

This Section explains the feature-based enhancement to the EDA module of EEND-EDA \cite{horiguchi2020end}, by learning a conversational summary representation. Figure \ref{fig:network} presents the proposed network architecture.

\subsection{Summary vector estimation}

The original encoder of EEND-EDA takes an input sequence $X$ and outputs embedding $E$. Inspired by how the special \texttt{[CLS]} token is used in BERT \cite{devlin2018bert}, we modify the encoder to additionally estimate a summary representation $\hat{\textbf{u}} \in \mathbb{R}^{D}$ of the original input sequence:

\begin{equation}
    \hat{\textbf{u}}, E = \text{enc}(X).
\end{equation}

$E$ is input to the rest of EEND-EDA as usual \cite{leung2021robust}. We replace the original input zero vectors to $h^{\text{dec}}$ in (\ref{eq:orig_lstmd}), such that each estimated speaker-wise attractor is conditioned with $\hat{\textbf{u}}$:

\begin{align}
    \textbf{h}^{\text{dec}}_{s}, \textbf{c}^{\text{dec}}_{s} &= h^{\text{dec}}(\hat{\textbf{u}}, \textbf{h}^{\text{dec}}_{s-1}, \textbf{c}^{\text{dec}}_{s}) \qquad \qquad (s=1, ..., S).
\end{align}

At each iteration the LSTM decoder now sees a conversational summary representation of the input audio sequence. $h^{\text{enc}}$ remains unchanged from the original design, as seen in equation (16) of \cite{horiguchi2021encoder}.

\subsection{Summary vector formulation} \label{sec:summ-vec-form}

We outline three different methods to initialize conversational summary representation $\textbf{u}$ of the encoder:

\begin{itemize}
  \item Average pooling is used to summarize the mean presence of features in $X$.
  \item Max pooling is used to summarize the most activated presence of features in $X$.
  \item A randomly initialized parameter of size $D$ is added to the encoder network so that it's gradients can be updated by the training objective. 
\end{itemize}

\subsection{Conformer encoder modification}

Summary vector $\textbf{u}$ is concatenated with input feature sequence $X$ for input to modules of the Conformer encoder \cite{gulati2020conformer}, such that the first frame of every sequence corresponds to a special summary representation.

The dashed box on the right of Figure \ref{fig:network} shows our modification to the Conformer encoder model architecture when using an input with a special summary representation. The structure of the encoder block is the same, using two feed-forward modules (FFN) in the style of Macaron-Net \cite{lu2019understanding}, a Multi-Head Self Attention Module (MHSA), Convolution Module (Conv) and layer normalization (LayerNorm). However, the hidden summary vector skips the Convolution module to ensure it maintains a global representation. The hidden feature sequence passes through the Convolution module with a residual connection. The hidden summary vector is concatenated back with the hidden feature sequence before input to the second feed-forward module.

Mathematically, the outputs $\hat{\textbf{u}}, E$ of a single modified Conformer block with summary vector $\textbf{u}$ and input $X$ is:

\begin{equation}
    \begin{aligned}
        \tilde{\textbf{u}}, \tilde{X} &= (\textbf{u} \oplus X) + \frac{1}{2}\text{FFN}(\textbf{u} \oplus X) \\
        \textbf{u}', X' &= (\tilde{\textbf{u}} \oplus \tilde{X}) + \text{MHSA}(\tilde{\textbf{u}} \oplus \tilde{X}) \\
        X'' &= X' + \text{Conv}(X') \\
        \hat{\textbf{u}}, E &= \text{LayerNorm}((\textbf{u}' \oplus X'') + \frac{1}{2}\text{FFN}(\textbf{u}'\oplus X'')), \\
    \end{aligned}
\end{equation}
where the $\oplus$ notation denotes the concatenation of tensors.

\section{Experiments} \label{sec:data-experiments}

\begin{table}[t]
\caption{Datasets used at various experimental stages.}
\label{tab:data}
\resizebox{\linewidth}{!}{%
\begin{tabular}{@{}llll@{}}
\toprule
\textbf{Stage} & \textbf{Dataset}                                                                                                                                         & \textbf{\#Speakers}                                                                  & \textbf{\#Mixtures}                                                      \\ \midrule
Pre-training 1 & LibriSpeech ($\beta$=2)                                                                                                                                  & 2                                                                                    & 100,000                                                                  \\ \midrule
Pre-training 2 & LibriSpeech ($\beta$=2, 2, 5, 9)                                                                                                                         & 1 - 4                                                                                & 400,000                                                                  \\ \midrule
Fine-tuning    & \begin{tabular}[c]{@{}l@{}}VoxConverse Dev. \\ DIHARD III Dev. (Train)\\ DIHARD III Dev. (Val.)\\ MagicData-RAMC Train\\ AMI Mix\\ AMI SDM1\end{tabular} & \begin{tabular}[c]{@{}l@{}}1 - 20\\ 1 - 10\\ 1 - 10\\ 2\\ 3 - 5\\ 3 - 5\end{tabular} & \begin{tabular}[c]{@{}l@{}}216\\ 162\\ 41\\ 289\\ 136\\ 135\end{tabular} \\ \midrule
Testing        & \begin{tabular}[c]{@{}l@{}}VoxConverse Evaluation Set\\ DIHARD III Evaluation  Set\\ MagicData-RAMC Test\\ AMI Mix\\ AMI SDM1\end{tabular}               & \begin{tabular}[c]{@{}l@{}}1 - 21\\ 1 - 9\\ 2\\ 3, 4\\ 3, 4\end{tabular}             & \begin{tabular}[c]{@{}l@{}}259\\ 232\\ 43\\ 16\\ 16\end{tabular}      \\ \bottomrule
\end{tabular}
}
\end{table}

\subsection{Data}

Table \ref{tab:data} presents the datasets used across all experiments for each stage of training and evaluation. LibriSpeech \cite{panayotov2015librispeech}\footnote{\url{https://www.openslr.org/12/}} recordings are used for both pre-training stages and are segmented using WebRTC VAD\footnote{\url{https://www.github.com/staplesinLA/denoising_DIHARD18/}}. The 960 hours LibriSpeech training set is used to simulate mixtures for 1, 2, 3 and 4 speakers by the protocol defined in \cite{horiguchi2020end}\footnote{\url{https://github.com/hitachi-speech/EEND/}}. Where $\beta$ is used to control the silence duration and overlap ratio. 

A number of public datasets covering a wide range of acoustic environments were used for fine-tuning. This included the VoxConverse \cite{chung2020spot}\footnote{\url{https://www.robots.ox.ac.uk/~vgg/data/ voxconverse/}} and DIHARD III \cite{ryant2020third, ryant2020thirdeval} Development datasets, MagicData-RAMC training set \cite{yang2022open} and the AMI Mix and SDM1 training sets \cite{carletta2006ami}. 

VoxConverse is a real world conversational dataset extracted from YouTube videos. Both the development set (20.3 hours) and evaluation set (43.5 hours) contains a significant proportion of overlapping speech. The DIHARD III dataset is a collection of various challenging datasets from 11 domains with different recording equipment and environments. The DIHARD III Development set, consisting of 203 mixtures, was split into training and validation sets by a ratio of 80\%:20\% per domain. Magic-RAMC is a high-quality Mandarin conversational speech dataset recorded on mobile phones by native speakers. The AMI Meeting Corpus contains 100 hours of recordings for close-talking (Mix) and far-field (SDM1) microphones in three rooms with varying acoustic properties in English with mostly non-native speakers.

The VoxConverse and DIHARD III Evaluation sets, Magic-RAMC test set and both AMI Mix and SDM1 test sets are used for evaluation. 

\subsection{Experimental setup}

The baseline model is EEND-EDA with no use of summary vector representations. For all models, the architecture of the encoder consisted of four stacked Conformer blocks ($N=4$) each with four attention heads. The Conformer encoder made use of Macaron style \cite{lu2019understanding} feed-forward layers with 1024 hidden units and outputted 256-dimensional frame-wise embeddings.  No positional embeddings were used. Inputs to the model were 23-dimensional log Mel-filterbanks with a frame length of 25 ms and a frame shift of 10 ms. Input features passed through a 10-fold convolutional subsampling module ($\text{Conv. Sub}$) consisting of two convolutional layers with kernel sizes $\{3, 5\}$ and stride $\{2, 5\}$. The convolutional upsampling module (\text{Conv. Up}) used two transposed convolutional layers with batch normalisation and ReLU activation. Here, kernel sizes $\{3, 5\}$, strides $\{2, 5\}$ and output padding $\{1, 0\}$ are used. We use the additive margin penalty on diarization results only and set $m$ to $0.35$ \cite{leung2021robust}.

Each model was pre-trained for two stages on the LibriSpeech simulated mixtures. In the first stage of pre-training each model was trained on mixtures containing only 2 speakers for 100 epochs. In the second stage each model was then trained for 25 epochs on a concatenation of mixtures containing 1 to 4 speakers. Models were adapted for a total of 500 epochs during the fine-tuning stage.
 
The Adam optimizer \cite{kingma2014adam} and Noam scheduler \cite{vaswani2017attention} with 100,000 warm-up steps was used during pre-training on simulated mixtures. At fine-tuning the Adam optimizer was used with a fix learning rate of $1 \times 10^{-5}$. Due to the time complexity of PIT, the model was trained to output the four most dominant speakers. Attractor existence loss $\mathcal{L}_{exist}$ was used to update only the EDA module by cutting the computational graph of all inputs to the EDA LSTM encoder to disable back propagation to the preceding layers.

For all stages of training a batch size of 64 was used. Unless specified, input recording lengths were set to 5000. Chunk shuffling was also used at fine-tuning \cite{leung2021end}. All models were trained on one GeForce RTX 3090 GPU for recording lengths of 5000. This included $\sim94$ hours for pre-training and $\sim22$ hours for fine-tuning. Up to four GPUs were used when increasing recording lengths to 20000 frames. Each model used 8.1M parameters. The evaluation metric is the DER with no collar tolerance. DER is composed of missed speech, false alarm and speaker error metrics \cite{anguera2012speaker}.

% Due to fixing the output number of speakers during training to four, results up to four speakers are \enquote{known} and recordings containing a higher number 

\section{Results} \label{sec:results}

\subsection{Results on summary vector initialisation}

% \begin{table}[]
% \caption{DER results for 1 to 4 speakers for all test datasets.}
% \label{tab:4-spk-avg}
% \centering
% \begin{tabular}{@{}llccc@{}}
% \toprule
% \textbf{Model} & \multicolumn{1}{c}{\textbf{MS}} & \textbf{FA} & \textbf{SPK-ERR} & \textbf{DER} \\ \midrule
% Baseline       & 7.93                            & 4.23        & 5.29             & 17.45        \\
% Summ. Vec.     & 7.33                            & 4.39        & 4.75             & 16.47        \\ \bottomrule
% \end{tabular}
% \end{table}

\begin{table}[]
\caption{DER (\%) results of different methods for initialising summary vector across all test datasets combined.}
\label{tab:all-spk-avg}
\centering
\resizebox{\linewidth}{!}{%
\begin{tabular}{@{}lccccc@{}}
\toprule
\textbf{Model} & \textbf{NS2}   & \textbf{NS3}   & \textbf{NS4}   & \textbf{NS2 to NS4} & \textbf{NS2 to NS9} \\ \midrule
Baseline       & 12.01          & 20.27          & 28.50          & 17.89               & 22.40               \\
SR-AvgPool     & \textbf{11.99} & 19.96          & 28.62          & 17.89               & 22.26               \\
SR-MaxPool     & 12.74          & 20.24          & 26.52          & 17.71               & 21.38               \\
SR-Learned     & 12.46          & \textbf{19.65} & \textbf{24.49} & \textbf{16.86}      & \textbf{20.50}      \\ \bottomrule
\end{tabular}
}
\end{table}

\begin{table}[]
\caption{DER (\%) results for each test set.}
\label{tab:all-datasets}
\centering
\resizebox{\linewidth}{!}{%
\begin{tabular}{@{}llcccc@{}}
\toprule
\textbf{Dataset}                & \textbf{Model} & \textbf{NS2} & \textbf{NS3} & \textbf{NS4} & \textbf{NS2 to NS4} \\ \midrule
\multirow{2}{*}{DIHARD III}     & Baseline       & 10.32        & 23.62        & 34.29        & \textbf{14.09}      \\
                                & SR-Learned     & 10.90        & 25.64        & 34.95        & 14.82               \\ \midrule
\multirow{2}{*}{VoxConverse}    & Baseline       & 9.23         & 19.60        & 26.15        & 16.59               \\
                                & SR-Learned     & 8.66         & 14.59        & 15.34        & \textbf{11.94}      \\ \midrule
\multirow{2}{*}{MagicData-RAMC} & Baseline       & 14.37        & -            & -            & \textbf{14.37}      \\
                                & SR-Learned     & 14.94        & -            & -            & 14.94               \\ \midrule
\multirow{2}{*}{AMI Mix}        & Baseline       & -            & 15.81        & 21.65        & 21.01               \\
                                & SR-Learned     & -            & 16.36        & 19.06        & \textbf{18.76}      \\ \midrule
\multirow{2}{*}{AMI SDM1}       & Baseline       & -            & 17.85        & 34.53        & 32.72               \\
                                & SR-Learned     & -            & 17.65        & 30.12        & \textbf{28.76}      \\ \bottomrule
\end{tabular}
}
\end{table}

The first experiment measured the DER performance of the baseline to three models that made use of conversational summary representations. This compared the different methods described in Section \ref{sec:summ-vec-form} to initialize the summary vector.

Table \ref{tab:all-spk-avg} presents the results across all test datasets combined. For recordings containing two to four active speakers (NS2 to NS4), we observe a relative DER improvement of 1.03 \% using learned summary representations (SR-Learned) when compared to the baseline. Most gains were made in recordings with four active speakers (NS4). Initialising summary vector by applying average pooling (SR-AvgPool) or max pooling (SR-MaxPool) to the subsampled input sequence shows comparable performance to the baseline. SR-Learned further improves the absolute DER by 1.90 \% when compared to the baseline for recordings with two to nine active speakers (NS2 to NS9).

Table \ref{tab:all-datasets} shows a breakdown of results for each test dataset. On average, SR-Learned outperforms the baseline on VoxConverse and both AMI test sets, whilst performing comparably with the baseline on other test sets. The most gains can be observed in recordings with four active speakers.

\subsection{Results on varying input recording lengths}

\begin{table}[]
\caption{DER (\%) results for all test datasets combined when varying input recording lengths during fine-tuning.}
\label{tab:seconds}
\centering
\resizebox{\linewidth}{!}{%
\begin{tabular}{@{}lccccc@{}}
\toprule
\textbf{Model}  & \multicolumn{1}{l}{\textbf{NS2}} & \multicolumn{1}{l}{\textbf{NS3}} & \textbf{NS4}   & \textbf{NS2 to NS4} & \textbf{NS2 to NS9} \\ \midrule
Baseline 50s    & \textbf{12.01}                   & 20.27                            & 28.50          & 17.89               & 22.40               \\
Baseline 100s   & 12.97                            & 19.39                            & 25.62          & 17.49               & 21.33               \\
Baseline 150s   & 13.03                            & \textbf{16.97}                   & 25.79          & 17.36               & 21.22               \\
Baseline 200s   & 13.11                            & 23.67                            & 25.02          & 17.77               & 21.50               \\
SR-Learned 50s  & 12.46                            & 19.65                            & 24.49          & 16.86               & 20.50               \\
SR-Learned 100s & 12.35                            & 18.99                            & \textbf{22.24} & 16.03               & \textbf{19.74}      \\
SR-Learned 150s & 11.28                            & 19.02                            & 23.76          & 15.87               & 20.08               \\
SR-Learned 200s & 11.87                            & 18.59                            & 22.38          & \textbf{15.75}      & 20.02               \\ \bottomrule
\end{tabular}
}
\end{table}

By increasing the input recording lengths during fine-tuning the model was able to train on sequences containing a higher number of active speakers, subsequently increasing the proportion of speakers the EDA module was exposed to.

Table \ref{tab:seconds} shows the results for the baseline and SR-Learned models when varying input recording lengths during fine-tuning. Results for recording lengths of 5000 frames (50s) are copied from Table \ref{tab:all-datasets}. Both models achieve lower a DER for three and four active speakers, whilst the baseline degrades in performance for recordings with two active speakers. The best performing baseline uses recording lengths of 15000 frames (150s). SR-Learned improves this with an absolute DER improvement of 1.49 \% for two to four active speakers. SR-Learned yields greater improvements when increasing the length of input recordings, showing an absolute DER improvement of 1.11 \% for two to four active speakers when compared to the result for 50s. Improvements for two to nine active speakers could be negligible due to the model only being trained to diarize up to four active speakers.

\subsection{Insights on the behavior of EDA}

\begin{table}[]
\caption{Dot-product similarity of output embeddings and attractors, split by ground-truth labels. Where labels 1 - 4 refer to frames where individual speakers are active. Labels 5 and 6 denote overlapping and silent frames, respectively.}
\label{tab:dot}
\centering
\resizebox{\linewidth}{!}{%
\begin{tabular}{lrrrrr}
\multicolumn{6}{c}{(a) Baseline}                                                                                                                                                        \\ \hline
                                                               & \multicolumn{1}{l|}{}  & \multicolumn{4}{c}{\textbf{Attractors}}                                                       \\
                                                               & \multicolumn{1}{l|}{}  & \multicolumn{1}{c}{1} & \multicolumn{1}{c}{2} & \multicolumn{1}{c}{3} & \multicolumn{1}{c}{4} \\ \hline
\multirow{6}{*}{\begin{sideways}\textbf{Labels}\end{sideways}\hspace{-5pt}} & \multicolumn{1}{r|}{1} & \textbf{5.4}          & -6.6                  & -4.2                  & -5.8                  \\
                                                               & \multicolumn{1}{r|}{2} & -6.6                  & \textbf{3.2}          & -3.7                  & -5.0                  \\
                                                               & \multicolumn{1}{r|}{3} & 1.7                   & -4.9                  & \textbf{-1.8}         & -4.1                  \\
                                                               & \multicolumn{1}{r|}{4} & -5.7                  & -1.8                  & -5.0                  & \textbf{1.4}          \\
                                                               & \multicolumn{1}{r|}{5} & 0.9                   & -0.2                  & -2.1                  & -0.3                  \\
                                                               & \multicolumn{1}{r|}{6} & -23.9                 & -13.1                 & -12.3                 & -14.9                 \\ \hline
\end{tabular}
\hspace{5pt}

\begin{tabular}{lrrrrr}
\multicolumn{6}{c}{(b) SR-Learned}                                                                                                                                                      \\ \hline
                                                               & \multicolumn{1}{l|}{}  & \multicolumn{4}{c}{\textbf{Attractors}}                                                       \\
                                                               & \multicolumn{1}{l|}{}  & \multicolumn{1}{c}{1} & \multicolumn{1}{c}{2} & \multicolumn{1}{c}{3} & \multicolumn{1}{c}{4} \\ \hline
\multirow{6}{*}{\begin{sideways}\textbf{Labels}\end{sideways}\hspace{-5pt}} & \multicolumn{1}{r|}{1} & \textbf{6.7}          & -9.2                  & -6.6                  & -7.5                  \\
                                                               & \multicolumn{1}{r|}{2} & -9.5                  & \textbf{8.9}          & -8.0                  & -7.4                  \\
                                                               & \multicolumn{1}{r|}{3} & -6.6                  & -7.4                  & \textbf{5.0}          & -7.3                  \\
                                                               & \multicolumn{1}{r|}{4} & -9.9                  & -10.2                 & -7.9                  & \textbf{8.3}          \\
                                                               & \multicolumn{1}{r|}{5} & -1.9                  & -1.7                  & -1.4                  & 0.7                   \\
                                                               & \multicolumn{1}{r|}{6} & -18.5                 & -27.2                 & -14.5                 & -10.0                 \\ \hline
\end{tabular}
}
\end{table}

To gain more understanding about the behaviour of EDA, we investigate the dot-product similarity between the output up-sampled embeddings and attractors for both the baseline and SR-Learned models. Table \ref{tab:dot} presents the results for a four speaker mixture from the AMI Mix evaluation set. The speaker frames, overlapping frames and silent frames were assigned based on the ground-truth labels. Attractors for both models are seen to be well separated from embedding frames related to silence. SR-Learned attractors are shown to be more separated from frames not relating to their represented speaker when compared to the baseline.

\section{Conclusion} \label{sec:conclusion}

In this work, we introduced conversational summary representations for end-to-end neural diarization with encoder-decoder based attractor calculation. Of the three methods investigated for initializing the summary vector, it was found that an additional parameter to the encoder network yielded the most gains. Further improvements were observed when increasing input recording lengths. Adopting the learned summary representation, the model demonstrated an absolute DER performance improvement of 1.90 \% (for recording lengths of 5000 frames) for two to nine active speakers when compared to the baseline.

\bibliographystyle{IEEEtran}
\bibliography{mybib}

\end{document}